\documentclass[twocolumn]{aastex63}
\usepackage{multirow}
\usepackage{savesym}
\savesymbol{tablenum}
\usepackage{siunitx}
\restoresymbol{SIX}{tablenum}
\usepackage{longtable}
\usepackage{color}
\usepackage{chemformula}

\newcommand{\msun}{$M_\odot$}
\newcommand{\lov}{$L_A$}
\newcommand{\mov}{$M_A$}

\submitjournal{ApJ}

\shorttitle{Stellar models for Betelgeuse}
\graphicspath{{./}{figures/}}
\begin{document}

\title{Stellar models of Betelgeuse constrained using observed surface conditions}

\correspondingauthor{Luo Tianyin}
\email{tianyin@g.ecc.u-tokyo.ac.jp}

\author{Luo Tianyin}
\affiliation{Department of Astronomy, Graduate School of Science, The University of Tokyo, 7-3-1 Hongo, Bunkyo City, Tokyo 113-0033, Japan}% \\

\author{Umeda Hideyuki}
\affiliation{Department of Astronomy, Graduate School of Science, The University of Tokyo, 7-3-1 Hongo, Bunkyo City, Tokyo 113-0033, Japan}% \\

\author{Yoshida Takashi}
\affiliation{Yukawa Institute for Theoretical Physics, Kyoto University, Kitashirakawa Oiwakecho, Sakyo-ku, Kyoto 606-8502, Japan}

\author{Takahashi Koh}
\affiliation{Max Planck Institute for Gravitational Physics, D-14476 Potsdam, Germany}

\begin{abstract}

We study stellar models for Betelgeuse using the HR diagram and surface abundances as observational constraints.
Previous studies on Betelgeuse have not systematically investigated the surface abundances, but we believe they can be impacted by, and thus be used as an observational constraint for various parameters such as initial mass, rotation, and overshoot scheme. 
We investigate stellar models with varying initial mass as they evolve past the main sequence, and we examine the red supergiant (RSG) properties in detail. 
For each mass, we vary the initial rotation up to $\sim \SI{300}{km\:s^{-1}}$, and test two different overshoot parameters.
Overall, the acceptable initial mass range is 12 to 25 \msun{}, but for non-rotating models only, the range is decreased to 15 to 24 \msun{}.
Also for rotating models, we find that $v/v_{\rm K} = 0.3$ is the upper limit for initial rotation, as more rapidly rotating models are unable to fit to Betelgeuse's surface abundances as an RSG.
In addition, we report two possibilities for the current stage of evolution, core helium burning or core carbon burning and beyond. 
We find that certain 17 \msun{} models could fit to both stages.
Finally, we discuss the implications of our results in the context of merger scenarios which have been suggested as a mechanism to attain the observed surface velocity of Betelgeuse.

\end{abstract}

\keywords{supergiants --- stars:rotation --- stars:evolution --- 
stars:individual (Betelgeuse)}

\section{Introduction} \label{sec:intro}

Betelgeuse (also known as $\alpha$ Orionis) is one of the brightest M-type supergiants in the night sky.
Due to its brightness and relative proximity, it has long been a popular target of observations, and especially being used as an archetype to study the properties of red supergiants (RSG).
Multi-band observations of Betelgeuse have been plenty \citep{Wilson1992, Burns1997, Uitenbroek1998}, and its periodic variable behaviour has also been well documented in the past decades \citep{Goldberg1984, Smith2009}. 

In late 2019, the star underwent a well publicised dimming episode over several months \citep{Guinan2019,Guinan2020}, followed by an equally puzzling rapid rise in luminosity in 2020 \citep{Sigismondi2020}.
This dimming brought its brightness to levels below what is typically expected from its inherent variability \citep{Levesque2020, Harper2020b}, generating speculation among both academics and the general public alike. 
There have been various suggestions to explain the dimming, such as conjectures which suggest an imminent supernova event, and other less exciting proposals where a shroud of dust that entered into our line of sight \citep{Gupta2020, Harper2020a} or changes occurred in the star's photosphere \citep{Dharmawardena2020}.

Unfourtunately, due to the difficulty in obtaining precise distance measurements \citep{Harper2008, Harper2017}, many of Betelgeuse's fundamental stellar properties remain uncertain. 
As a result, there is currently no clear consensus on Betelgeuse's evolution history, and thus we cannot immediately explain the dimming episode nor predict its future course of evolution with only observations.
With that taken into consideration, scientists have began taking a different approach in the past several years with the aid of powerful computers. 
By calculating stellar models and then comparing their properties with Betelgeuse's observable properties, we can constrain the numerous variable parameters which impact stellar evolution \citep{Meynet2013, Dolan2016, Wheeler2017, Nance2018, Joyce2020}. 
The eventual goal is to construct a complete stellar model which conforms to Betelgeuse, and this will not only allow us to understand Betelgeuse's past and future, but also elucidate the detailed inner mechanisms of RSGs as a whole.

In particular, \citet{Dolan2016} had calculated a grid of non-rotating models and found results which favoured a progenitor mass of $20^{+5}_{-3} M_\odot$ that is ascending the red giant branch. 
Their choice of observable constraints included the luminosity, surface temperature, mass loss rate, and radius.
Surface abundances were also discussed, but were only used for their best fit 20 \msun{} model, for which they found an adequate match with observed values.
Also, they had suggested that initial rotation could potentially allow less massive stars to satisfy Betelgeuse's constraints, but that was not something they investigated in their study. 
Another study by \citet{Wheeler2017}, which was the first of a series of three papers called the Betelgeuse Project \citep{Wheeler2017, Nance2018, Sullivan2020}, examined both non rotating and rotating models in the 15 to 25 \msun{} range.
However, their main purpose was to study the surface rotation of Betelgeuse, so they did not apply any observable constraints except the surface rotation and the HR diagram (albeit with rather large error bars).
They found that regardless of initial mass, their models only produced fits for the surface rotation near the base of the red giant branch, and that rotation cannot be maintained at a satisfactory level as the star continues to evolve past that point. 

Another recent study by \citet{Joyce2020} took a different approach.
They used asteroseismic simulations in addition to hydrodynamical calculations, and their observational parameter of choice was Betelgeuse's pulsation periods, which included a $\approx$ 400 day cycle identified as the fundamental frequency, and a $\approx$ 185 day cycle which was described as the first overtone.
Through the examination of their models' pulsation patterns, they were able to determine a best fit initial mass value of 18 to 21 \msun{}, which is slightly stricter than \citet{Dolan2016}.
In addition, they were able to derive new radius and distance estimates, which were in good agreement with the values measured by \textit{Hipparcos}.

Clearly, there are still some conflicting results regarding Betelgeuse's progenitor model and past evolution history, and this study is motivated by the prospect of filling in the gap in information left by the aforementioned studies.
In this study, we investigate stellar models of a range of initial masses, including both non rotating and rotating models, with a focus on the use of surface carbon, nitrogen, and oxygen (CNO) abundances as the observational constraint as the star evolves as an RSG.
Up until now, the surface abundances have largely been neglected as a constraint parameter, as studies have not done an in depth investigation over a large number of models.
The aim is to find the times during which a model can be a fit for Betelgeuse in order to identify viable progenitor model properties.
And with the use of surface abundances, stellar parameters involved in the mixing process, such as the rotational velocity and overshoot parameter, can be focused on in particular.
In the process, the mystery of Betelgeuse's current stage of evolution, which includes discussion relating to the recent dimming episode that spurred speculation about a possible supernova, will also hopefully become better understood.

This text will be organised as follows. In section \ref{sec:methods}, the theoretical model and parameters and the choice of observational constraints will be explained. 
In section \ref{sec:results}, the results of our calculations will be presented. 
In section \ref{sec:discussion}, the results of this study will be discussed in the context of contemporary literature.
Finally in section \ref{sec:summary}, an overall summary will be provided.

\section{Methods} \label{sec:methods}

\subsection{Observational Constraints} \label{subsec:obscons_method}

For ease of comparison, observational constraints for the HR diagram used in this paper will be the same as those adopted by \citet{Dolan2016}, namely $\log L/L_\odot = 5.1 \pm 0.22$ and $T_{eff} = 3500 \pm \SI{200}{K}$, where $L$ and $T_{eff}$ are luminosity and effective temperature.
The adopted surface temperature is the result of an average of past studies as the surface temperature of Betelgeuse is known to vary, and the luminosity is derived from the distance measurement given by \citet{Harper2008}. 
A new distance measurement was reported by \citet{Harper2017}, but since it only differs from the 2008 results by 0.7$\sigma$, we stick with the 2008 results for ease of comparison. 
These same observational constraints are also used by \citet{Wheeler2017}, although with three times the uncertainty.

Regarding the surface abundances, we adopt the observed abundances of CNO elements relative to hydrogen given in \citet{Lambert1984} as the constraint. 
However, when considering the observed values, it is important to consider its dependence on the surface temperature because \citet{Lambert1984} reported varying relative abundances in the range of 3600 to \SI{3800}{K}. 
\citet{Dolan2016} argued that such a correlation in surface abundance and temperature could be neglected due to the inherent variability of Betelgeuse, and used the relative abundances at \SI{3800}{K} as their constraint. 
However, we believe it is more suitable to use the values given for \SI{3600}{K}, because the difference between the observed abundances at 3600 and \SI{3800}{K} is a non-negligible amount. 
\citet{Lambert1984} reports an error of $\pm 0.15$ (units are in dex) for each element at $3800 \pm \SI{100}{K}$, but we take the error range for our adopted surface abundance to be the difference between the values reported for 3600 and \SI{3800}{K} (see figure 6 in their paper). 
This difference was 0.12 for carbon, and 0.25 for nitrogen and oxygen, but for the case of carbon, we have decided it is more sensible to use the larger 0.15 value. 
Thus, our constraints for the relative abundances are $\epsilon_{\ch{C}} = 8.29 \pm 0.15$,
$\epsilon_{\ch{N}} = 8.37 \pm 0.25$, and $\epsilon_{\ch{O}} = 8.52 \pm 0.25$, 
where $\epsilon_i = \log(X_i/X_{\ch{H}}A_i) + 12$, $X_i$ is the mass fraction of element $i$,
$X_{\ch{H}}$ is the Hydrogen mass fraction, and $A_i$ is the average mass number of element $i$.

In addition, we also look at the ratio of N/O as another constraint, as it is a rather robust constraint that barely varies with respect to the surface temperature. From our adopted values, we find the logarithmic value of the N/O ratio is -0.15 (ie. the difference between $\epsilon_{\ch N} - \epsilon_{\ch O}$), and the error range is $\sim 0.05$ as reported in \citet{Lambert1984}.

\subsection{Model Description}

For this study, we use the 1D stellar evolution code HOSHI, which has been in continuous development by \citet{Takahashi2013, Takahashi2014, Takahashi2018a, Yoshida2019}. 
Stellar models are evolved from the zero-age main sequence (ZAMS) and terminated when the central temperature reaches $\log T_c = 9.2$, approximately corresponding to the period between core carbon and neon burning. 
We follow the nuclear burning using the nuclear reaction network of 300 species of nuclei \citep{Takahashi2018b}.
Nuclear reaction rates are taken from the JINA reaclib database v1 \citep{Cyburt2010}, except for the $^{12}C(\alpha,\gamma)^{16}O$ rate which is taken to be 1.2 times the value given in \citet{CF88}.
Evolution beyond this point is typically less than a few years and we assume the red giant branch properties would not be strongly affected.  
We initially investigate 15, 17, 20, and 25 \msun{} models, but we also add other models from 12 to 26 \msun{} as necessary in order to determine the upper and lower limits for initial mass. 
We consider the initial rotation and overshoot parameters as variables which can strongly affect the stellar evolution and the structure of that star.  

\subsubsection{Initial rotation}

The initial rotation is a parameter which is expected to have a large impact on the surface CNO abundances, as it applies a centrifugal effect as well as a meridian circulation effect to the star. 
In particular, the meridian circulation is a convective process where material is brought towards the surface along the axis of rotation, and flow towards the core occurs along the equatorial plane, and it plays a big role in the transportation of chemical elements to the surface \citep{Huang2004}.
In addition, rotation is known to favour convective mixing processes instead of inhibiting them \citep{Maeder2008}, so we also expect a larger initial rotation to result in more drastic changes in the model's surface abundances.
When compared to non rotating models, rotation would also result in an increase of the core size, thus enhancing production of CNO elements in the core regions, which would then be brought to the surface as the star evolves, through both the aforementioned convective process as well as during the dredge-up phase.
These factors all combine to impact the surface abundances of the star as a red supergiant.

In HOSHI, angular momentum transport and the chemical mixing process induced by rotation are taken with a diffusive treatment. The included rotation effect is described in detail in \citet{Takahashi2014}.
The initial surface velocity is prescribed using the Kepler velocity, $v_{\rm K} \equiv \sqrt{GM/R}$, where $G$ is the gravitational constant and $R$ is the stellar radius. 
The code then applies this velocity in the form of rigid body rotation for ZAMS models.
This should be considered a reasonable and sufficient assumption as the post-ZAMS evolution does not strongly depend on the star's formation history pre-ZAMS \citep{Hammerlae2013}. 
For initial rotation values, we have chosen $v/v_{\rm K} = 0.1$, 0.2, and 0.4, and the exact velocity values when applied to 15, 17, and 20 \msun{} models can be seen in table \ref{tab:surfvel}.
These values were taken based off the report from \citet{Georgy2012} and \cite{Ekstrom2012} which found a critical velocity of approximately $v/v_{crit} = 0.4$ to be the average initial rotational velocity based on the observed main sequence width on the HR diagram and the population of RSGs in our galaxy.
Here it should be noted that there is a slight discrepancy between $v_{crit}$ and $v_{\rm K}$, by a factor of $\sqrt{2/3}$. 
Indeed, when the values in table \ref{tab:surfvel} are compared with \citet{Ekstrom2012}, we find that our velocities are approximately 9\% to 19\% percent higher depending on the mass of the model.
Nevertheless, these initial rotation values covers the range of values currently accepted to be characteristic of massive stars, and will be adequate for the purpose of this investigation.
 
\begin{table}
    \caption{Initial surface velocity}
    \label{tab:surfvel}
    \begin{tabular}{ccc}
    \tableline
    Mass (\msun)   & $v/v_{\rm K}$ &  $v$ (km s$^{-1}$) \\
    \tableline
    \multirow{3}{*}{15}  & 0.1 & 79.0  \\
        & 0.2 & 157.6 \\
        & 0.4 & 296.1 \\\cline{1-3}
    \multirow{3}{*}{17}  & 0.1 & 81.0  \\
        & 0.2 & 161.2 \\
        & 0.4 & 315.1 \\\cline{1-3}
    \multirow{3}{*}{20}  & 0.1 & 83.9  \\
        & 0.2 & 166.7 \\
        & 0.4 & 328.9 \\
    \tableline
    \end{tabular}
\end{table}

\subsubsection{Overshoot and mixing}

We also investigate the effects of varying the convective overshoot parameter. 
This parameter has implications on the mass of the helium core, which in turn will dictate the advanced stage evolution.
In addition, convective flows are very efficient at mixing material, and is able to transport enriched material from the core up to the photosphere. 
For these reasons, we want to investigate its impact on the surface abundances of our stellar models as an RSG, in the context of Betelgeuse's observed properties.
In particular, this parameter governs the physics at the core-envelope boundary, and describes a diffusive process where material in the convective core ``overshoots" the boundary and mixes into the envelope.
In HOSHI, this process is described in equation form for the diffusion coefficient: $D = D_0 \exp{\left(\frac{-2\Delta r}{f_{ov}H_{\rm P}}\right)}$, where $D_0$ is the diffusion constant at the boundary, $\Delta r$ is the distance from the boundary, $f_{ov}$ is the overshoot parameter which can be varied, and $H_{\rm P}$ is the pressure scale height.

In short, we test two values for the overshoot parameter $f_{ov}$ for the main sequence, 0.03 and 0.01, but $f_{ov}$ is held constant at 0.002 after the core helium burning (characterized by central temperature $\log T_{\rm C} \ge 8.7$).
We believe this is enough to make a probe into the impact of the overshoot parameter, as the majority of the impact on the surface abundances should result from core activity during the longer lasting main sequence evolution. 
In this study, the naming conventions in \citet{Yoshida2019} will be followed, and the two overshoot models will be refered to as \lov{} ($f_{ov} =$ 0.03) and \mov{} ($f_{ov}=$ 0.01). These names stem from the fact those values were calibrated against early B-type stars in the Large Magellanic Cloud \citep{Brott2011} and AB stars in open clusters of the Milky Way \citep{Maeder1989, Georgy2012}, respectively.

\subsection{Other Parameters and Variables}

We have chosen to ignore several other variables found in previous studies for a variety of reasons.
First, we choose not to use the radius as an observational constraint due to its dependence on the highly uncertain distance measurement, as well its redundancy with the luminosity. 
However, we do discuss the radii of our models in context with contemporary literature in section \ref{sec:discussion}.

Furthermore, the mass loss rate applied to our models on the red giant branch is from \citet{deJager1988}, and is not varied among our models. 
According to \citet{Dolan2016}, who examined various mass loss rate parameterizations, the \citet{deJager1988} rate is larger than their adopted observational rate of 2 $\pm$ \SI{1e-6}{M_\odot yr^{-1}}, but they also note their adopted value can only be considered a lower limit, and realistic modelling of mass loss is complicated. 
Also, a recent study by \citet{Mauron2011} found the measured mass loss rate of several galactic RSGs agree well with the \citet{deJager1988} perscription.
Thus, we believe it is sufficient to use the \citet{deJager1988} mass loss rate for the purpose of this study.

Finally, in regards to the metallicity of Betelgeuse and RSGs, previous studies report a wide range of values. 
\citet{Ramirez2000} results show an [Fe/H] range of $0.05 \pm 0.14$, while \citet{Lambert1984}, on the basis of \citet{Luck1977, Luck1979}, suggests an enhanced [Fe/H] value of up to $\sim$ 0.2 is possible.
Here [Fe/H] = $\epsilon_{\ch{Fe}} - \epsilon_{\ch{Fe}, sol}$ represents relative abundance to solar values, and can be regarded as an overall indicator of metal abundance.
In this study, the main elements of our focus are carbon, nitrogen, and oxygen, so we can look at the [CNO/H] values as a benchmark.
From section \ref{subsec:obscons_method}, we can calculate the total [CNO/H] of our adopted values at \SI{3600}{K} to be 8.88. 
Our code defaults to the solar metallicity given in \citet{Asplund2009}, which also gives [CNO/H] = 8.88.
Thus, we believe the solar metallicity models are sufficient for the purpose of this study.

\section{Results} \label{sec:results}

We have evolved models ranging from 12 to 26 \msun{}, with varying initial rotation values. 
As an overview, table \ref{tab:solsumbrief} shows a summary of all the models which were calculated, and whether or not they provided a good fit to Betelgeuse during their evolution, and table \ref{tab:solsum} shows the timing of the fit for models labelled with $\circ$.
For simplicity, we devise a naming scheme to identify each model, in the form \textit{mmrrO}, where \textit{mm} is a two digit number for the initial mass, \textit{rr} refers to the initial rotation (\textit{no} is non-rotating, \textit{01} is $v/v_{\rm K} = 0.1$ and so forth), and finally \textit{O}, representing the overshoot parameter, is either \textit{M} or \textit{L}. 
For example, \textit{15noM} would refer to a 15 \msun{}, non-rotating model with \mov{} overshoot.

\begin{table*}[t]
    \centering
    \caption{Summary of the fit to Betelgeuse for all models which were tested}
    \label{tab:solsumbrief}
    %\resizebox{\columnwidth}{!}{%
    \begin{tabular}{c|c|c|c|c|c|c|c|c|c|c|c|c|c|c|c|c}
    \tableline
    Rotation & \multicolumn{2}{c|}{12 \msun}  & \multicolumn{2}{c|}{13 \msun} & \multicolumn{2}{c|}{15 \msun}  & \multicolumn{2}{c|}{17 \msun}   &  \multicolumn{2}{c|}{20 \msun} & \multicolumn{2}{c|}{24 \msun} & \multicolumn{2}{c|}{25 \msun} & \multicolumn{2}{c}{26 \msun} \\\cline{2-17}
    ($v/v_{\rm K}$) & \lov{} & \mov{} & \lov{} & \mov{} & \lov{} & \mov{}  & \lov{} & \mov{} & \lov{} & \mov{} & \lov{} & \mov{} & \lov{} & \mov{} & \lov{} & \mov{} \\
    \tableline
    no-rot & $\times$ & $\times$ & $\times$ & $\times$ & $\times$ & $\circ$ & $\circ$ & $\circ$ & $\circ$ & $\circ$ & $\times$ & $\circ$ & $\times$ & $\times$ & $-$ & $-$ \\
    \tableline
    0.1   & $\times$ & $\times$ & $\times$ & $\times$ & $\circ$ & $\circ$ & $\circ$ & $\circ$ & $\circ$ & $\circ$ & $-$ & $-$ & $\times$ & $\times$ & $\times$ & $\times$ \\
    \tableline
    0.2   & $\circ$ & $\times$ & $\circ$ & $\times$ & $\circ$ & $\times$ & $\circ$ & $\circ$ & $\times$ & $\circ$ & $-$ & $-$ & $\times$ & $\circ$ & $\times$ & $\times$ \\
    \tableline
    0.3  & $-$ & $-$ & $-$ & $-$ & $\times$ & $\times$ & $\times$ & $\circ$ & $\times$ & $\times$ & $-$ & $-$ & $\times$ & $\times$ & $-$ & $-$ \\
    \tableline 
    0.4  & $-$ & $-$ & $-$ & $-$ & $\times$ & $\times$ & $\times$ & $\times$ & $\times$ & $\times$ & $-$ & $-$ & $\times$ & $\times$ & $-$ & $-$ \\
    \tableline
    \end{tabular}\par
    \bigskip
    {\textbf Notes} -- $\circ$ represents a model which had a good fit for Betelgeuse on the red giant branch, $\times$ represents a model which did not, and $-$ represents models which were not calculated.
\end{table*}

\begin{table*}[t]
    \centering
    \caption{Summary of the fit timings for all models marked with $\circ$ in table \ref{tab:solsumbrief}}
    \label{tab:solsum}
    \begin{tabular}{c|c|c|ccc}
    Mass (\msun)  & Overshoot & $v/v_{\rm K}$ & $t_{col,u}$ (yr) & $t_{col,l}$ (yr) & $t_{total}$ (yr)\\
    \tableline
    12 & \lov{} & 0.2 & 5.90E+03  & 0  & 5.90E+03  \\
    \tableline
    13 & \lov{} & 0.2 & 1.25E+04  & 0  & 1.25E+04  \\
    \tableline
    \multirow{6}{*}{15} & \multirow{4}{*}{\lov} & \multirow{2}{*}{0.1} & 9.81E+05  & 8.98E+05 & 1.00E+05\\
     &  &  & 2.17E+04  & 0        & 2.17E+04\\\cline{3-6}
     &  & \multirow{2}{*}{0.2} & 9.99E+05  & 8.96E+05 & 1.03E+05  \\
     &  &  & 2.18E+04  & 0        & 2.18E+04  \\\cline{2-6}
     & \multirow{2}{*}{\mov}  & no & 1.31E+04 & 0 & 1.31E+04 \\\cline{3-6}
     &  & 0.1  & 1.13E+04  & 5.96E+03   & 5.36E+03 \\
    \tableline
    \multirow{8}{*}{17} & \multirow{5}{*}{\lov} & \multirow{2}{*}{no$^\dagger$} & 8.42E+05  & 6.57E+05  & 1.84E+05 \\
     &  &     & 1.04E+05  & 0  & 1.04E+05 \\\cline{3-6}
     &  & 0.1 & 8.38E+05  & 0  & 8.38E+05 \\\cline{3-6}
     &  & 0.2$^\dagger$ & 8.38E+05  & 6.14E+05  & 2.24E+05 \\\cline{2-6}
     & \multirow{4}{*}{\mov}  & no  & 2.15E+04 & 1.71E+04 & 4.41E+03 \\\cline{3-6}
     &  & 0.1 & 2.69E+04  & 1.87E+04 &  9.21E+03 \\\cline{3-6}
     &  & 0.2 & 9.29E+04  & 4.93E+04 &  4.36E+04 \\\cline{3-6}
     &  & 0.3 & 8.60E+05  & 8.53E+05 &  6.88E+03 \\
    \tableline
    \multirow{6}{*}{20}  & \multirow{2}{*}{\lov} & no$^\dagger$ & 8.66E+04  & 4.95E+04 & 3.71E+04 \\\cline{3-6}
     &  & 0.1  & 7.30E+05  & 9.70E+03  & 7.20E+05 \\\cline{2-6}
     & \multirow{3}{*}{\mov} & no & 3.47E+04 & 2.64E+04 & 8.21E+03 \\\cline{3-6}
     &  & 0.1  & 7.52E+04  & 1.85E+04 & 5.67E+04 \\\cline{3-6}
     &  & 0.2  & 1.76E+05  & 1.43E+05 & 3.29E+04 \\
    \tableline
    24 & \mov{} & no  & 8.10E+04  & 7.19E+04  & 9.07E+03  \\
    \tableline
    25 & \mov{}  & 0.2  & 2.92E+05  & 2.29E+05  & 6.27E+04   \\
    \tableline
    \end{tabular}\par
    \bigskip
    {\textbf Notes} -- $t_{total}$ refers to the total time of fit for that model. The time to collapse upper ($t_{col,u}$) and lower ($t_{col,l}$) limits refers to the time from the models' first and last time of good fit for Betelgeuse, respectively, until the end of the evolution. Models which had undergone a blue loop phase are marked with $\dagger$. Some models have more than one period of good fit, which are listed on separate lines in chronological order.
\end{table*}

In the rest of this section, we will provide detailed results on particular models. 
First, we will show the typical evolution of 15 \msun{}, non-rotating models as a reference. 
Following that, we will present the results of varying the other initial parameters.

\subsection{Non-rotating 15 \msun{} models}

\begin{figure}
    \centering
    \includegraphics[angle=-90, scale=0.34]{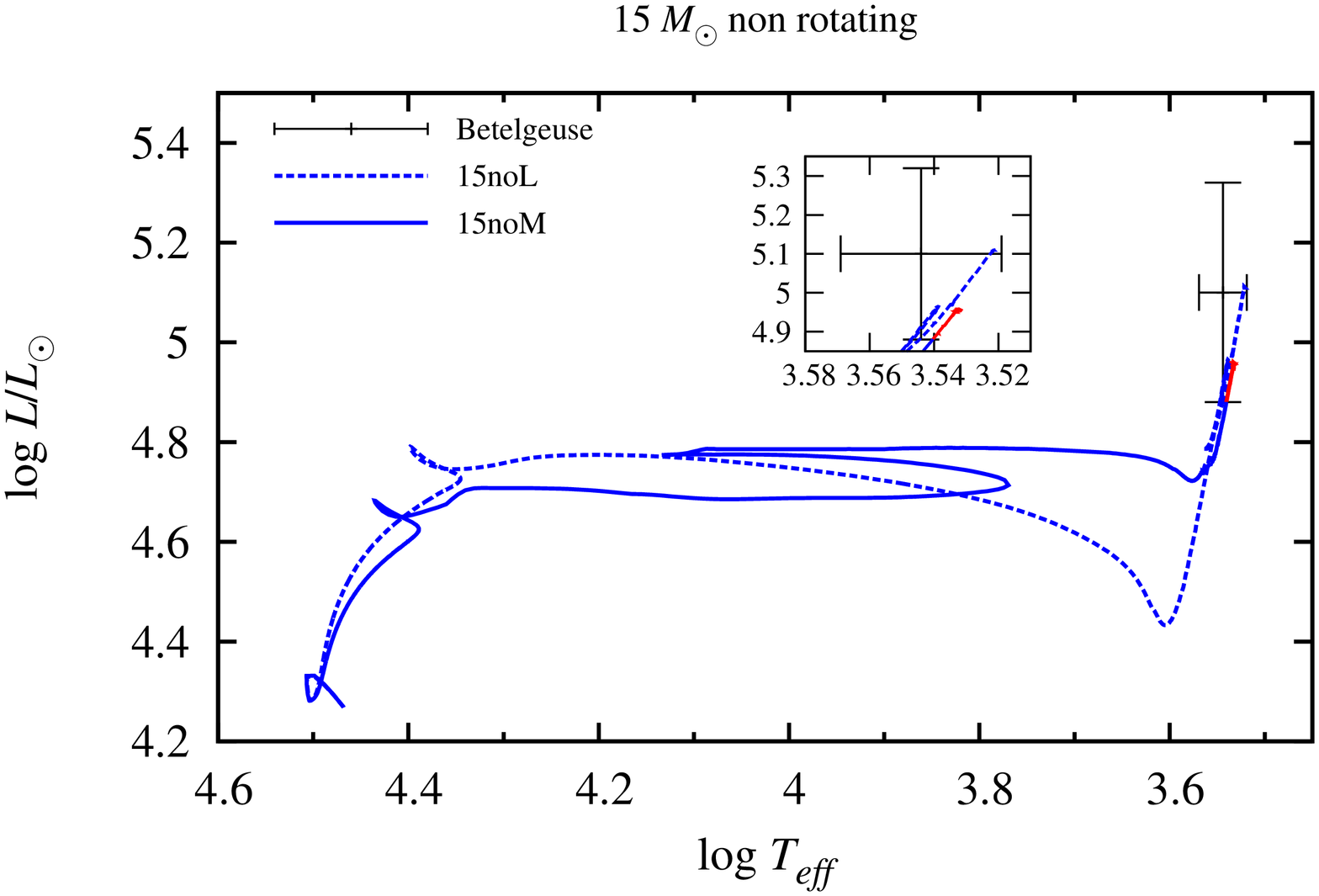}    \caption{Evolution tracks for the non-rotating 15 \msun{} models applied with different overshoot parameters. The inset plot is zoomed in and focused on the observed HR diagram position of Betelgeuse. The error bars are the same as those in \citet{Dolan2016}. The period where the model satisfies Betelgeuse's observational constraints are indicated in red.} 
    \label{fig:m15_norot}
\end{figure}

\begin{figure}
    \centering
    \includegraphics[angle=-90, scale=0.34]{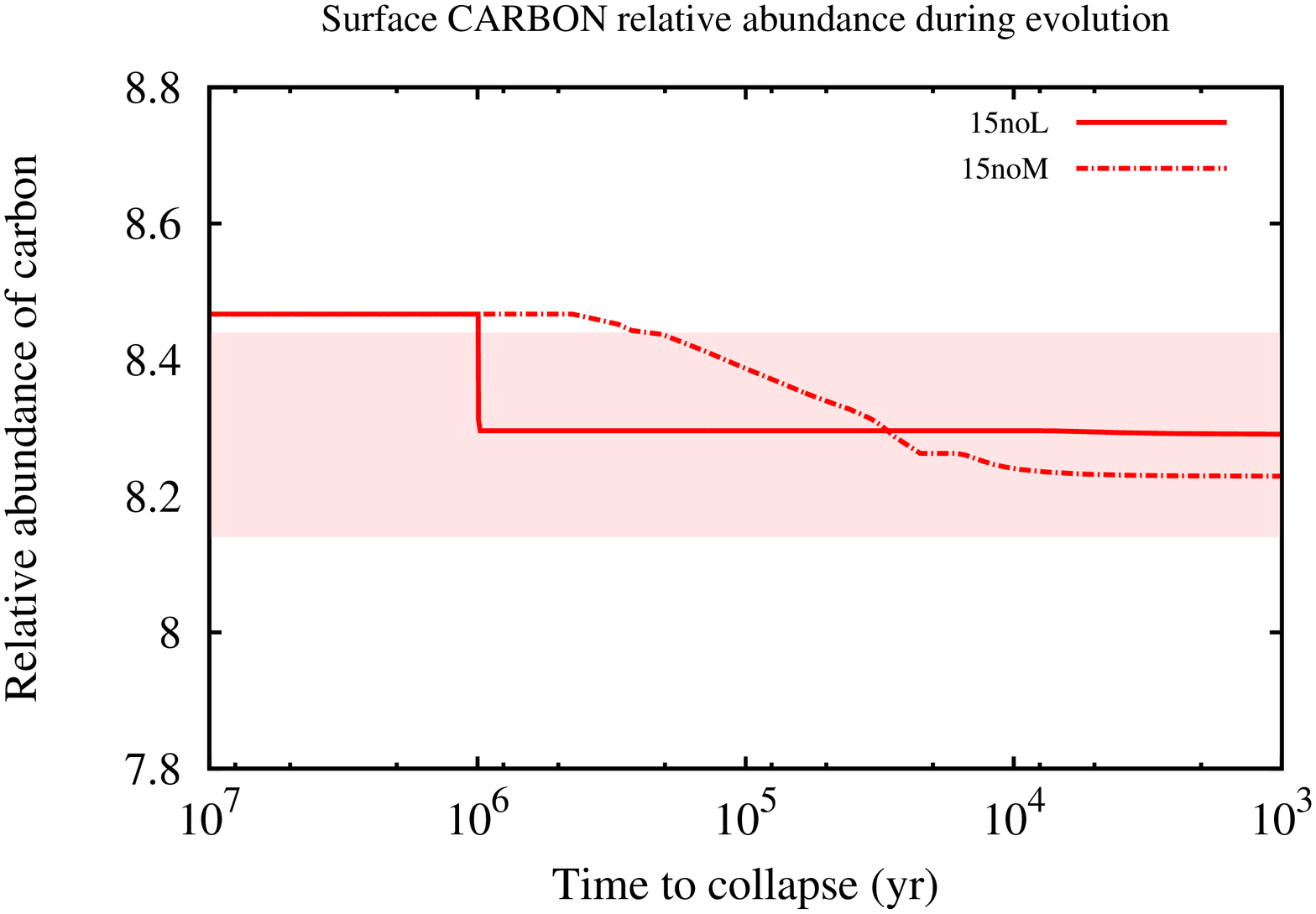}
    \includegraphics[angle=-90, scale=0.34]{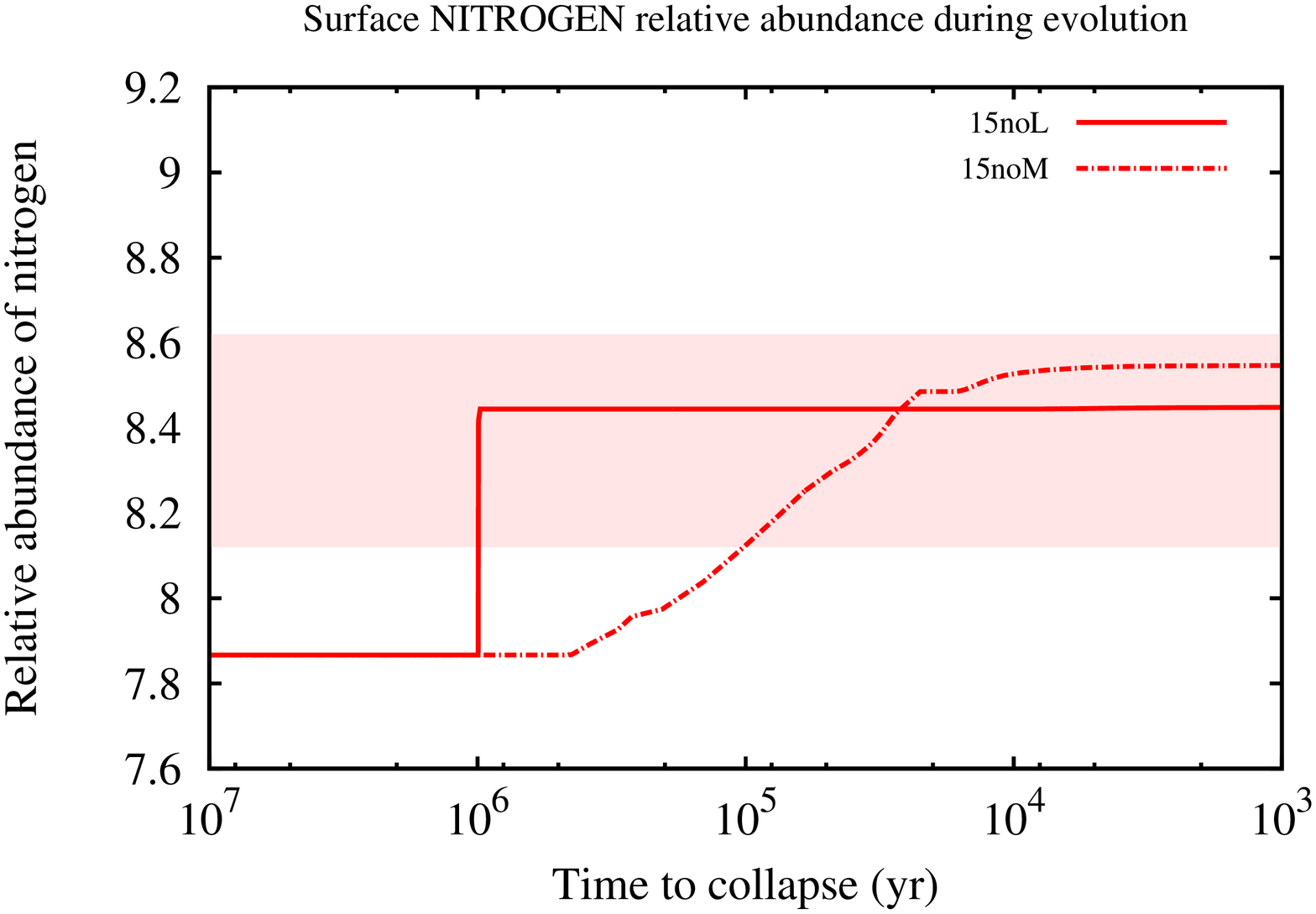}
    \includegraphics[angle=-90, scale=0.34]{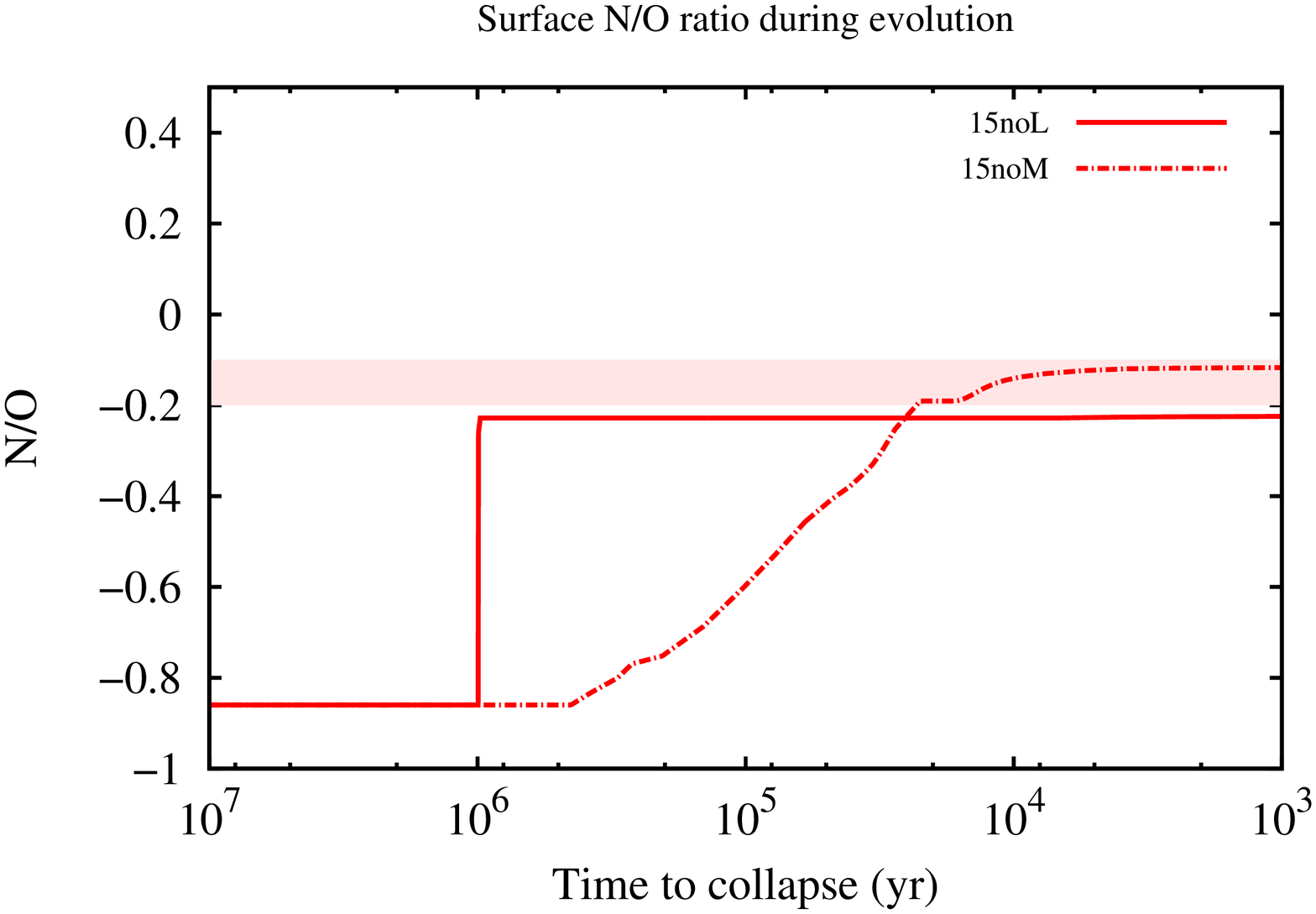}    \caption{Relative surface abundance of carbon, nitrogen, and the surface N/O ratio for 15 \msun{} models with no rotation. Top panel is carbon, middle panel nitrogen, and bottom panel the N/O ratio. ``Time to collapse'' refers to the time until the end of the evolution. The region shaded in red indicates the adopted observational constraints as given in section \ref{sec:methods}. Solid lines are \lov{} overshoot models, while dashed lines are \mov{} overshoot models.}
    \label{fig:norotsurf}
\end{figure}

Figure \ref{fig:m15_norot} shows the complete evolution of both \textit{15noM} and \textit{15noL} models. 
In the case of the \textit{15noM} model, the fit occurs during the very end of the evolution when the model is finally able to reach the observed luminosity error range.
This fit lasts until the end of the evolution, approximately $\sim\SI{e4}{yr}$, and covers the late core helium burning and core carbon burning stages.

On the other hand, the \textit{15noL} model is able to attain a high enough luminosity very early during its red giant branch evolution, before the star has even experienced its first dredge-up. 
The star then contracts and begins to dim, but it will slowly recover the luminosity after it begins core helium burning.

Figure \ref{fig:norotsurf} shows the change of surface abundances during the evolution.
The change in surface oxygen is insignificant, so we omit that plot, and instead, the surface N/O ratio is shown, which is a much stricter constraint on our results.
We can see that in the case of the higher $f_{ov} = 0.03$ overshoot \lov{} model, almost all of the change in surface abundances occurs during the dredge-up, and the abundances remain unchanged during subsequent red giant branch evolution.
As a result, the surface N/O ratio remains outside the error range and this model is not a good fit for Betelgeuse.
On the other hand, in the lower $f_{ov} = 0.01$ overshoot model, the surface CNO abundances are constantly changing during the red giant branch. 
This allows the model to become a good fit for Betelgeuse during the later stages of its evolution, including core helium and core carbon burning stages.

\subsection{Variation of initial rotation}

\begin{figure}
    \centering
    \includegraphics[angle=-90, scale=0.34]{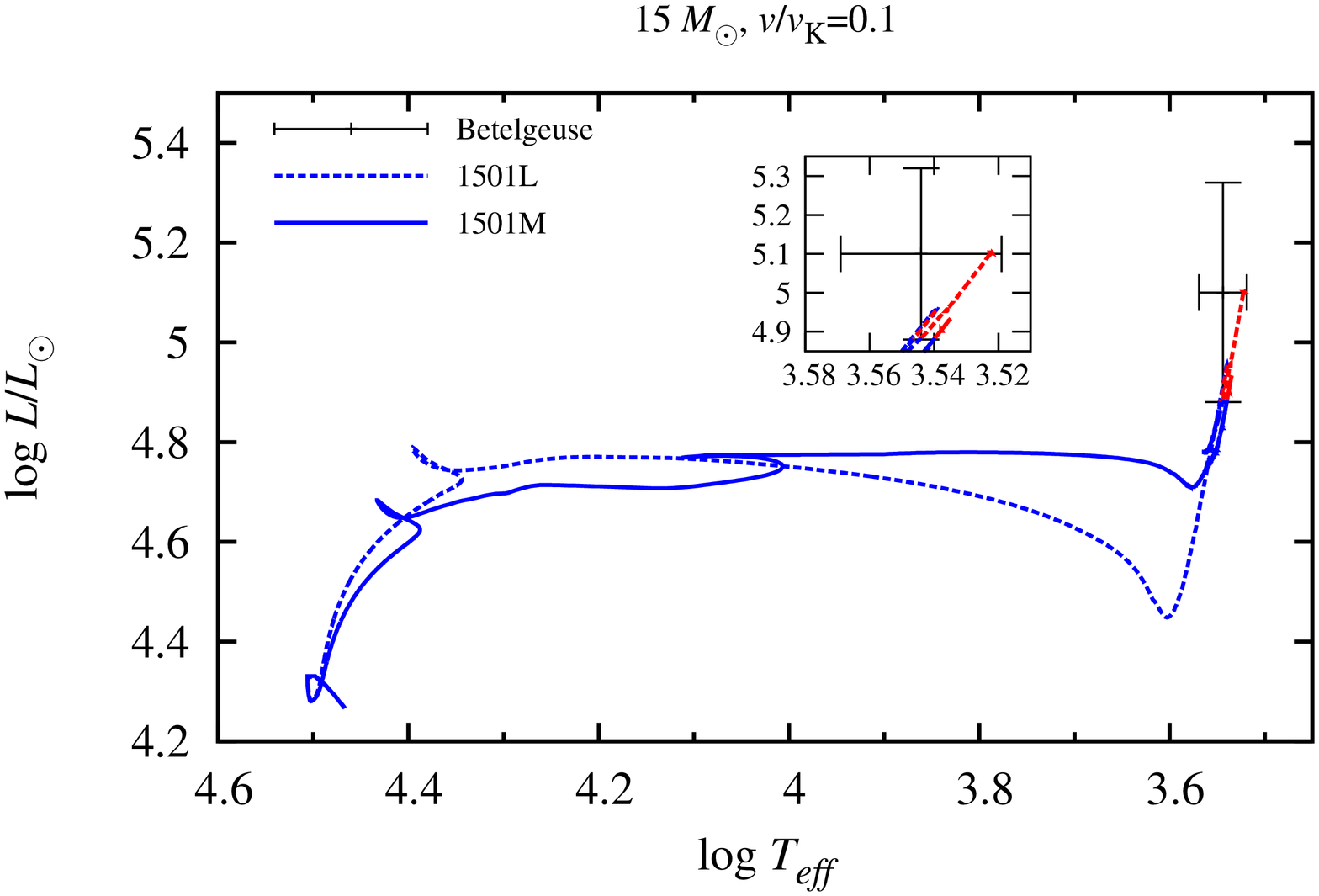}
    \includegraphics[angle=-90, scale=0.34]{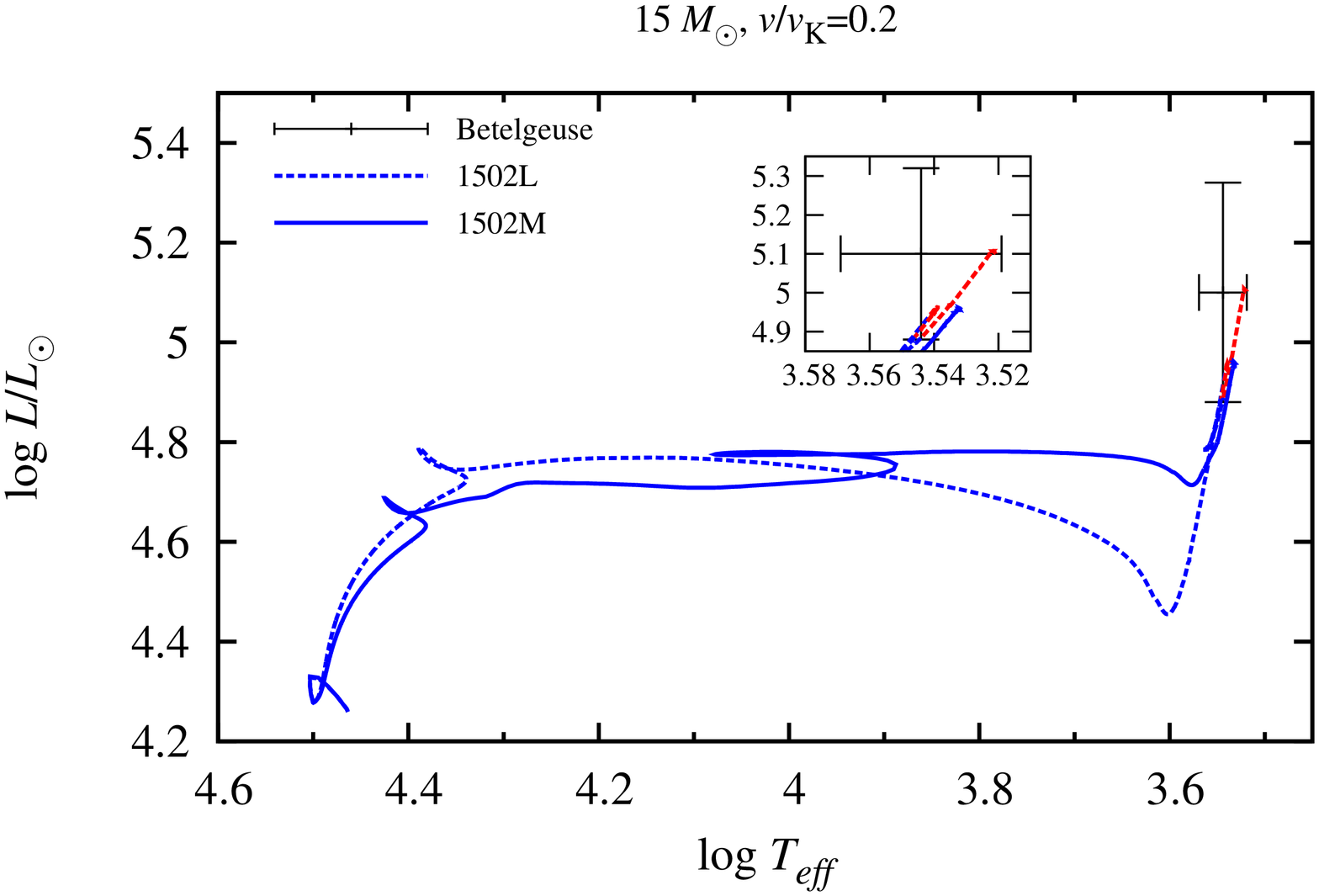}
    \caption{Evolution tracks for the 15 \msun{} rotating models applied with different overshoot parameters. The inset plot is zoomed in and focused on the observed HR diagram position of Betelgeuse. The error bars are the same as those in \citet{Dolan2016}. The period where the model satisfies Betelgeuse's observational constraints are indicated in red.} 
    \label{fig:m15rot}
\end{figure}

In figure \ref{fig:m15rot}, the HR diagrams of the $v_{\rm K} =$ 0.1 and 0.2 models are shown.
The impact on the HR diagram by varying initial rotation is minuscule, and their evolution tracks are nearly identical.

\begin{figure}
    \centering
    \includegraphics[angle=-90, scale=0.34]{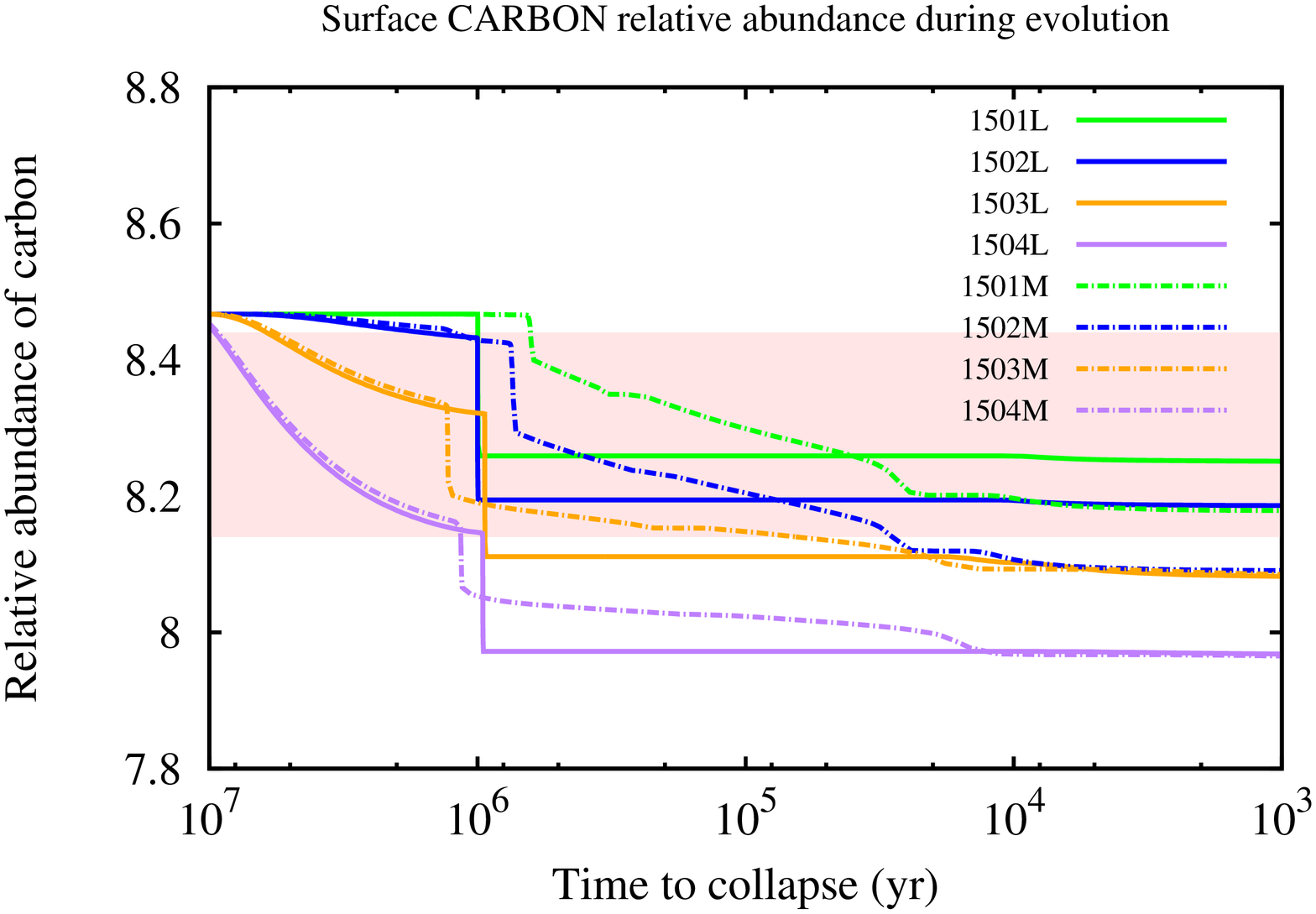}
    \includegraphics[angle=-90, scale=0.34]{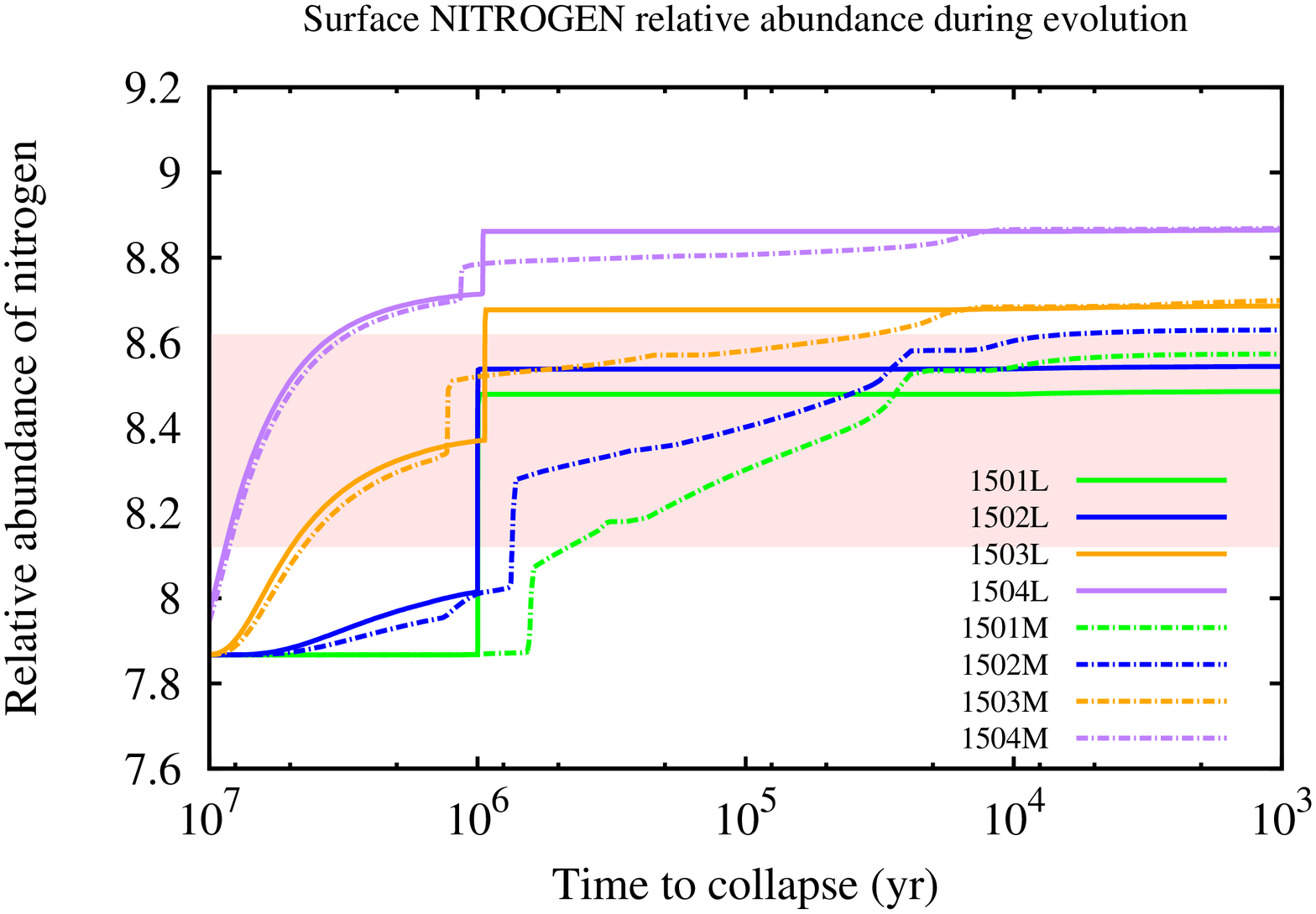}
    \includegraphics[angle=-90, scale=0.34]{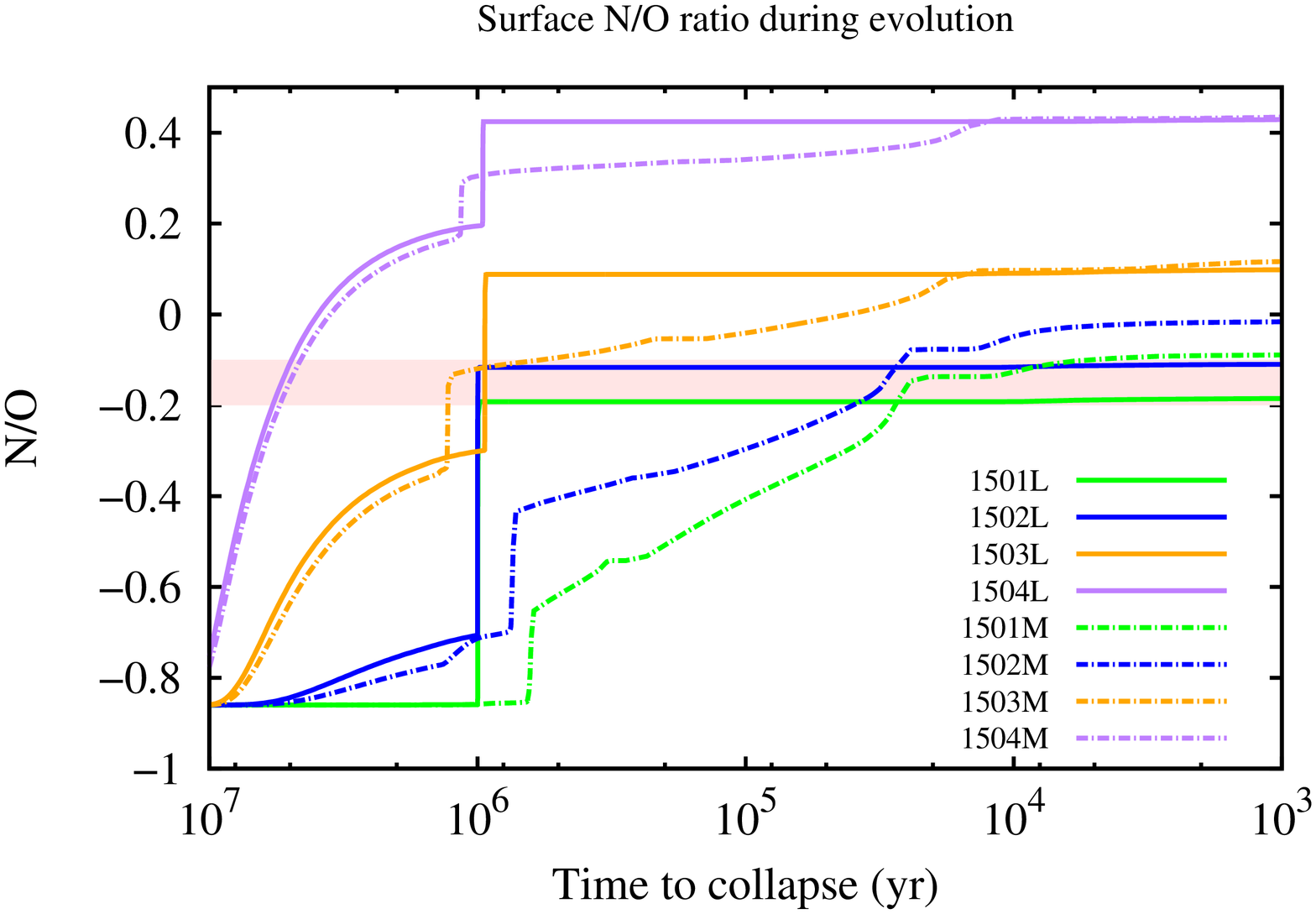}
    \caption{Same as figure \ref{fig:norotsurf}, but for 15 \msun{} models with varying initial rotation. }
    \label{fig:m15surf}
\end{figure}

Rather, the effect on the surface abundances is profound, and can be seen in figure \ref{fig:m15surf}.
The surface abundances are a limiting factor on the timing of the fit for Betelgeuse for these 15 \msun{} models.
For models of the same overshoot parameter, we can see the distinct monotonic relationship between the surface abundances of each CNO element and the initial rotation. 
For carbon and oxygen, higher initial rotation results in a lower abundance during the red giant branch evolution, while the reverse applies for surface Nitrogen. 
An increase in initial rotational velocity also leads to a change in surface abundance earlier during its evolution. 
This can be seen in models with $v/v_{\rm K} = 0.4$, where the onset of changes occurs almost at the beginning of the main sequence evolution.

As a result, only a specific range of initial rotation values allow for the model to reproduce a fit for all observational constraints. 
As mentioned before, in the case of non-rotating 15 \msun{} models, the \textit{15noL} model was unsatisfactory due to N/O ratio. 
However, referring back to table \ref{tab:solsumbrief}, if the initial rotation is increased to $v/v_{\rm K} = 0.1$ or 0.2, then we find the \textit{1501M, 1501L, and 1502L} models are all able to produce a fit to Betelgeuse. 
Both \lov{} models fit until the end of the evolution, while the \textit{1501M} model briefly fits for $\sim \SI{5e3}{yr}$ near the very end of the evolution. 
Models with initial rotation larger than $v/v_{\rm K} = 0.2$ also suffer from unsatisfactory N/O ratios, but in this case the surface nitrogen becomes much more abundant than surface oxygen. 
This upper limit of the acceptable initial rotation at $v/v_{\rm K} = 0.2$ (or 0.3 for 17 \msun{}) is noteworthy, since our chosen initial rotation velocities were based on \citet{Georgy2012, Ekstrom2012}, who had suggested that $v/v_{crit} = 0.4$ (or $v/v_{\rm K} \approx 0.33$) is a typical value to reproduce the RSG population in our galaxy.
This may suggest that the initial rotation velocity of Betelgeuse was slower than average
or the current prescription of rotation induced mixing has a problem. 
In either case, efforts to reproduce rotating models would be considered worthwhile.

\subsection{Variation of initial mass}

\begin{figure}
    \centering
    \includegraphics[angle=-90, scale = 0.34]{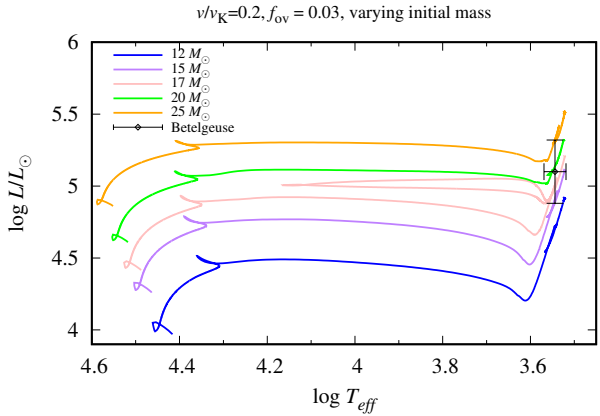}
    \caption{Evolution tracks for 12 to 25 \msun{} models with $v/v_{\rm K} = 0.2$ and \lov{}.}
    \label{fig:02rot}
\end{figure}

The initial mass of a model mainly affects the luminosity during the evolution, but also can slightly affect the surface abundances. 
From figure \ref{fig:02rot}, we can see that across models with the same initial rotation and overshoot parameter, the variation of luminosity is large, and becomes a limiting factor for many models at the extreme ends of the initial mass range.
Overall, for the suitable initial mass range of Betelgeuse progenitors, we find the lower initial mass limit at 12 \msun{} with the model \textit{1202L}.
This model is extremely limited by its luminosity, only being able to fit for $\sim \SI{6e3}{yr}$ at the very end of its evolution, indicating that any lower mass would not not viable. 
At the higher end of the initial mass range, we find a brief period of fit in the \textit{2502M} model (not shown in figure \ref{fig:02rot}, but it is indeed limited by its luminosity).
One noteworthy point is the overshoot parameter at both ends of the acceptable initial mass range, with the \lov{} overshoot required for the lower limit, and the \mov{} overshoot for the upper limit.
\citet{Dolan2016} had suggested that initial rotation would be key to allow lower mass models (ie. $\leq 15$ \msun{}) to fit to Betelgeuse, but our results suggest that an increase in the overshoot parameter is more crucial.
In the 17-20 \msun{} mass range, we see our models maintain much longer periods of fit, as they spend their entire red giant branch evolution within the observed HR diagram constraints, and both overshoot parameters produce viable models. 
This does corroborate with the results from \citet{Dolan2016} and \citet{Joyce2020} that the best fit for Betelgeuse's progenitor mass is in this range.

\begin{figure}
    \centering
    \includegraphics[angle=-90, scale=0.34]{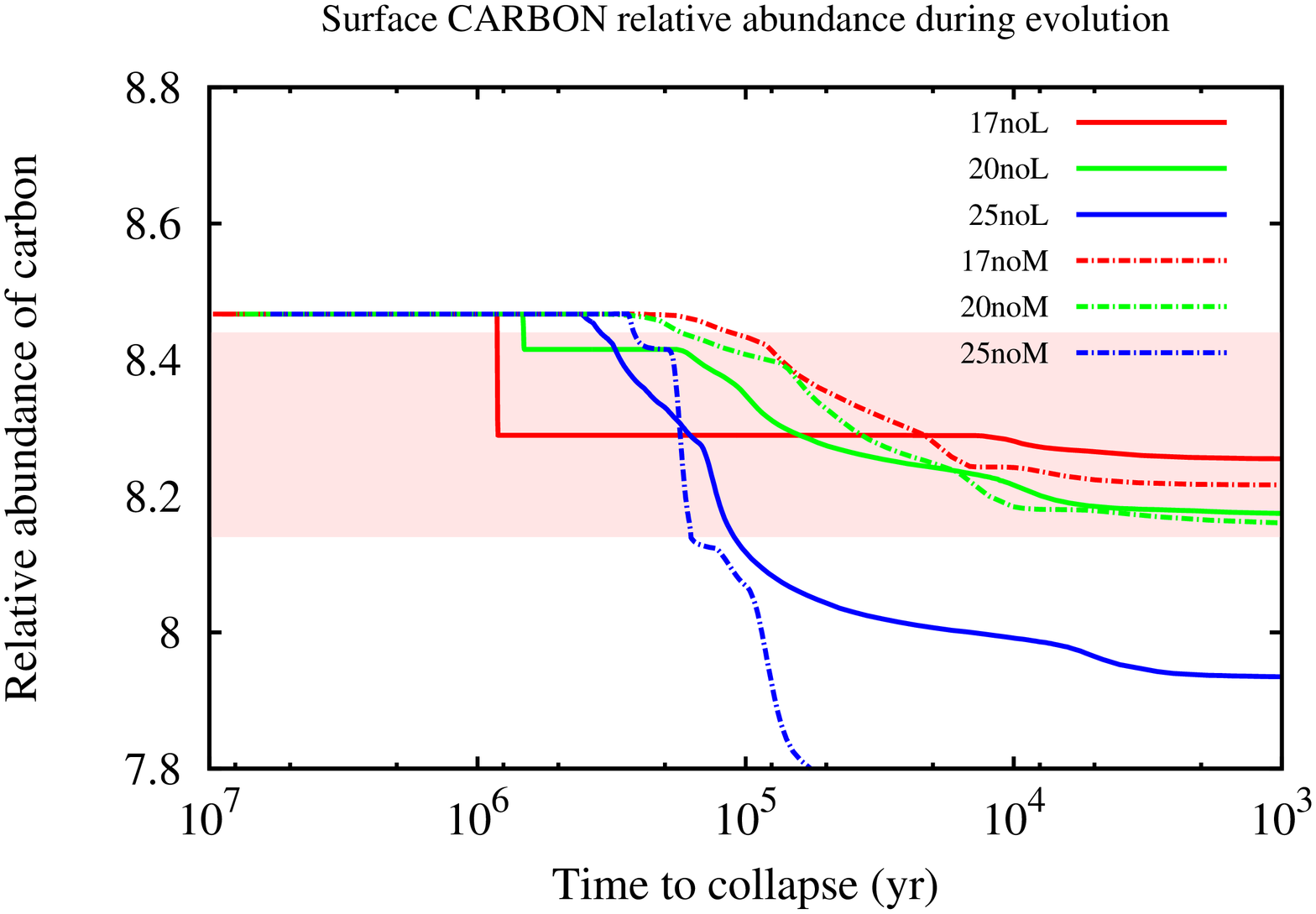}
    \includegraphics[angle=-90, scale=0.34]{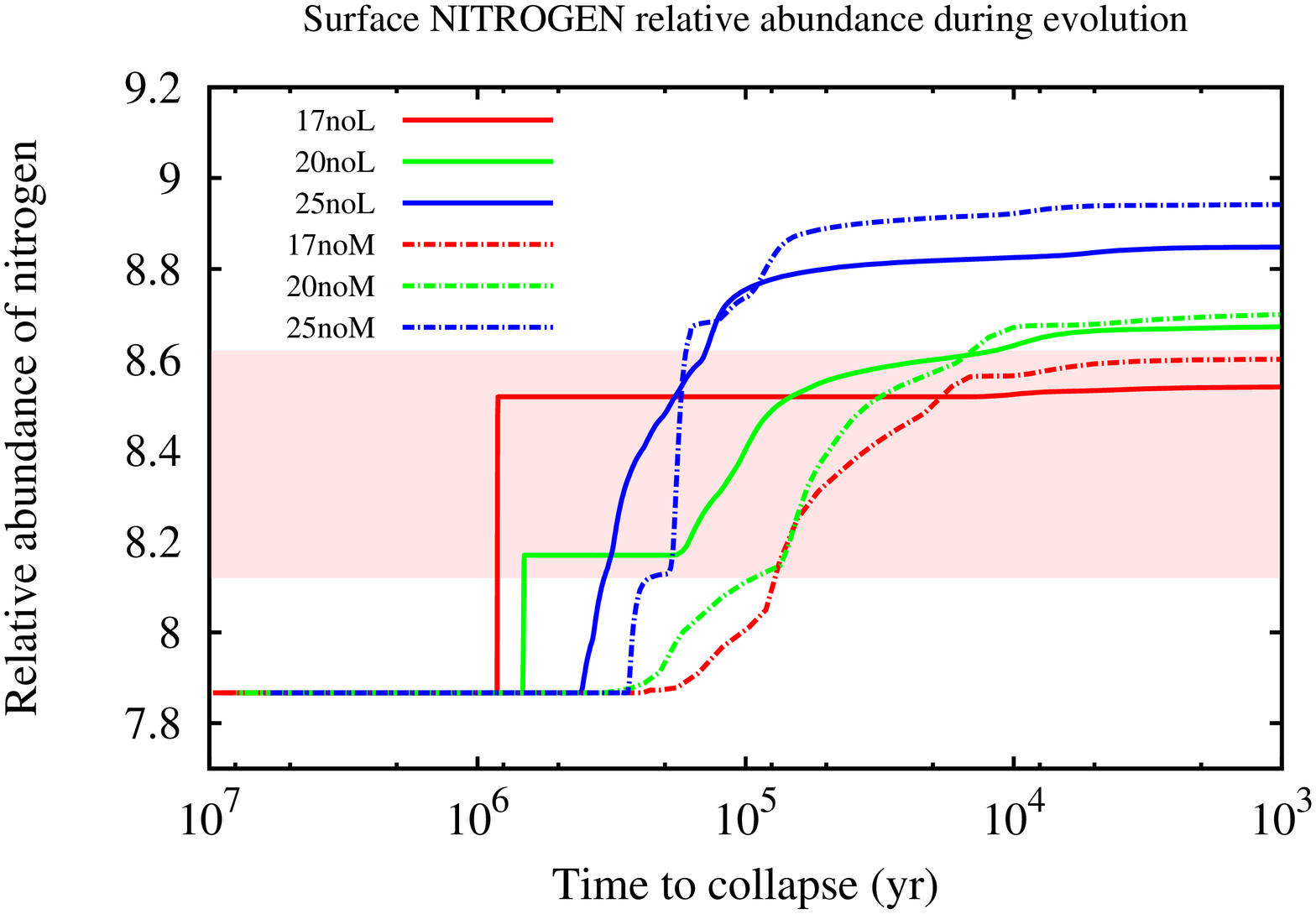}
    \includegraphics[angle=-90, scale=0.34]{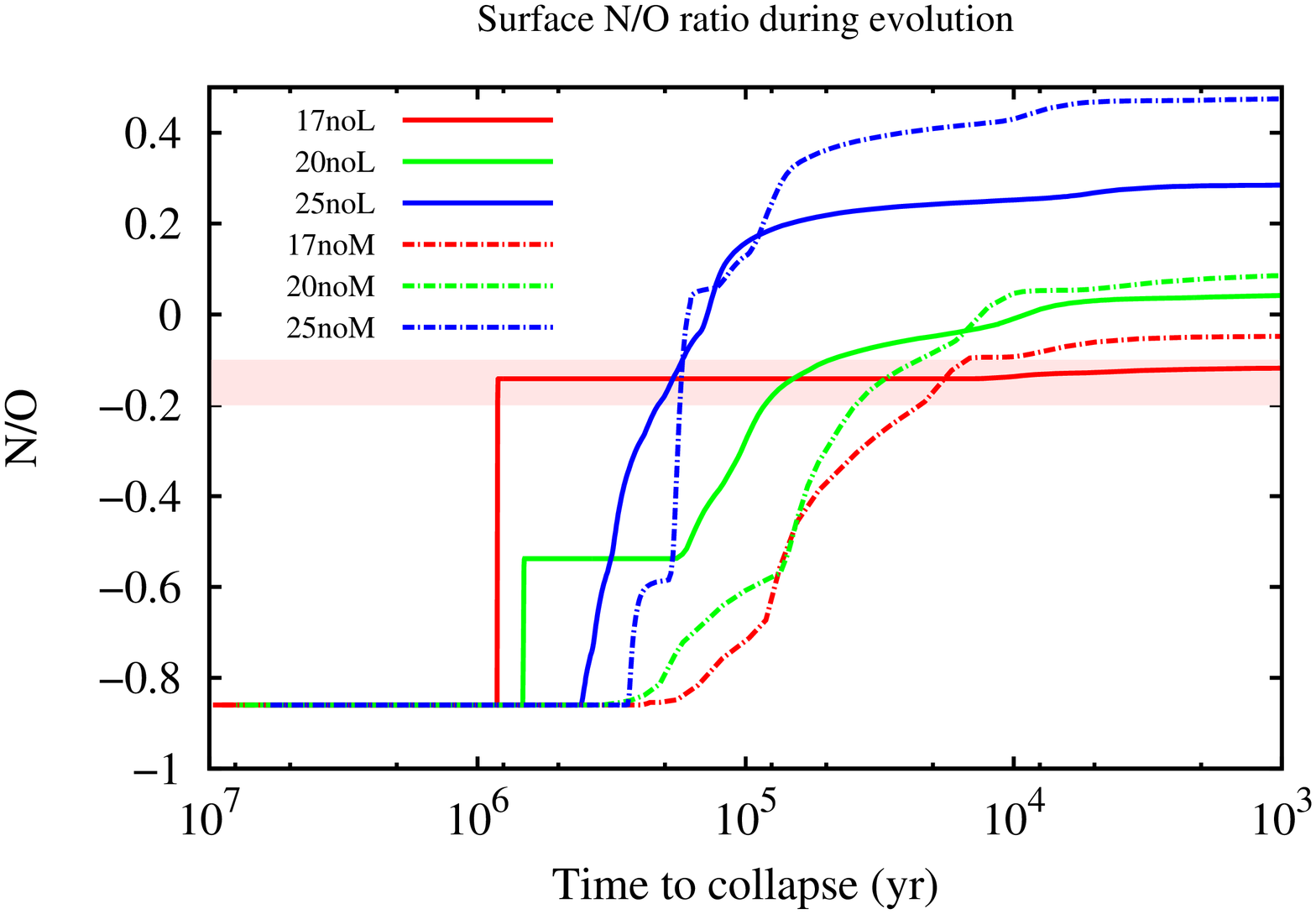}
    \caption{Same as figure \ref{fig:norotsurf}, but for 17, 20, and 25 \msun{} non-rotating models.}
    \label{fig:masssurf}
\end{figure}

The impact of the initial mass on surface abundances is much less dramatic, with the main differences arising in the N/O ratios. 
In figure \ref{fig:masssurf}, we can see that all models regardless of initial mass show no change in surface abundances before the dredge-up, allowing all models to have a time period where they can fit to Betelgeuse's observed surface abundances.
Also, larger masses have higher surface nitrogen and lower carbon and oxygen on the red giant branch, this allows the \textit{17noL} model to fit to Betelgeuse until the end of its evolution, as opposed to the \textit{15noL} model, which has an insufficient N/O ratio of $-0.23$.

Considering that the mass of Betelgeuse is not readily measurable, it is generally the most important parameter to be derived from these numerical simulation studies.
\citet{Dolan2016} favoured a best fit model with initial mass of $20^{+5}_{-3} M_\odot$, currently ascending the red giant branch during core helium burning.
They also suggested that initial rotation would be necessary for a lower mass model $\sim 15 M_\odot$ to satisfy the luminosity of Betelgeuse. 
\citet{Joyce2020}, using a different approach of combined asteroseismic and hydrodynamical simulations, reported a model derived initial mass of 18 to 21 \msun{}.
In our study, with the use of surface abundances as observational constraints, we find viable models for both non- and initially rotating models between 12 and 25 \msun{}, depending on the initial parameters, such as the overshoot parameter.

\section{Discussion} \label{sec:discussion}

\subsection{Current stage of evolution} \label{subsec:tcol_disc}

\begin{figure}
    \centering
    \includegraphics[angle=-90, scale=0.34]{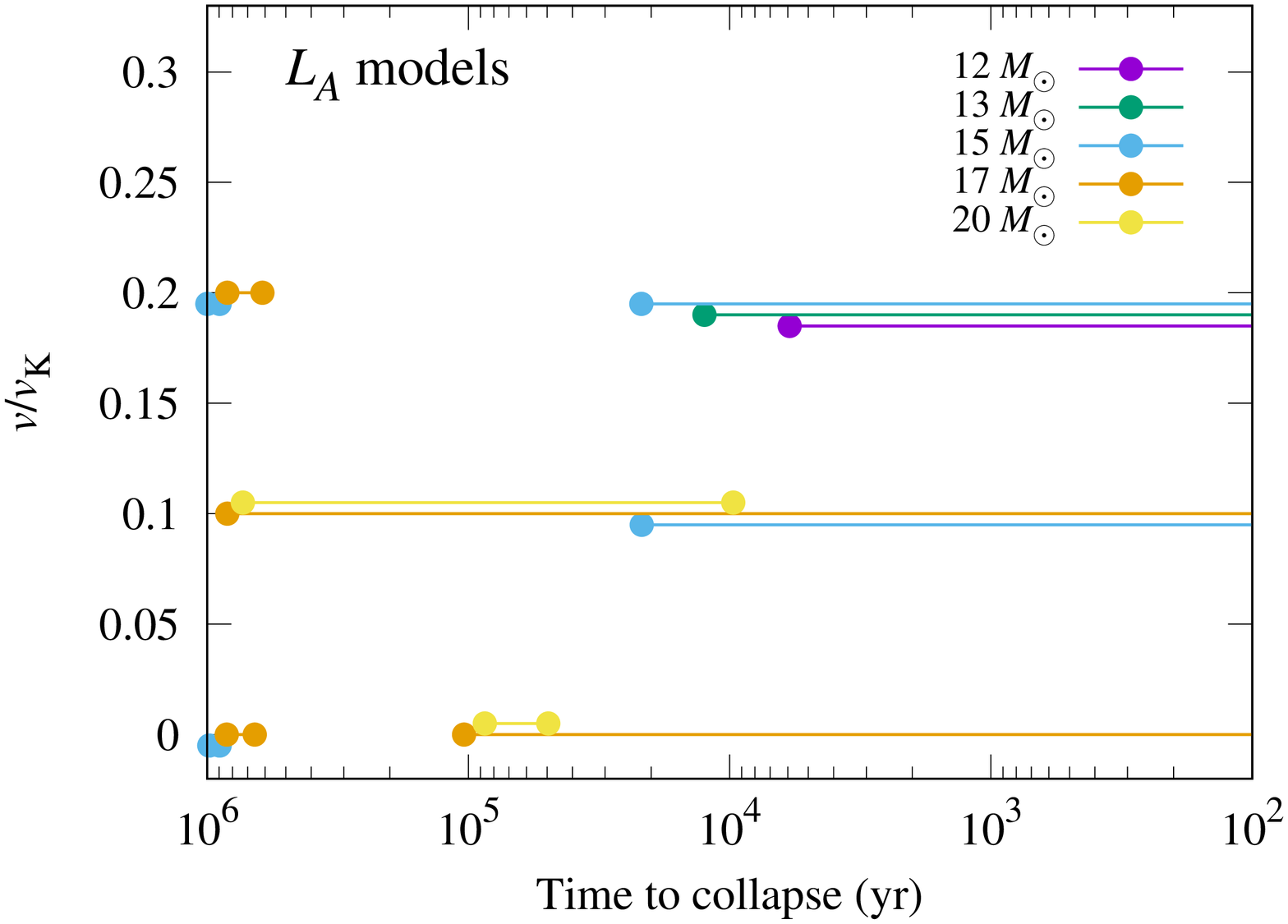}
    \includegraphics[angle=-90, scale=0.34]{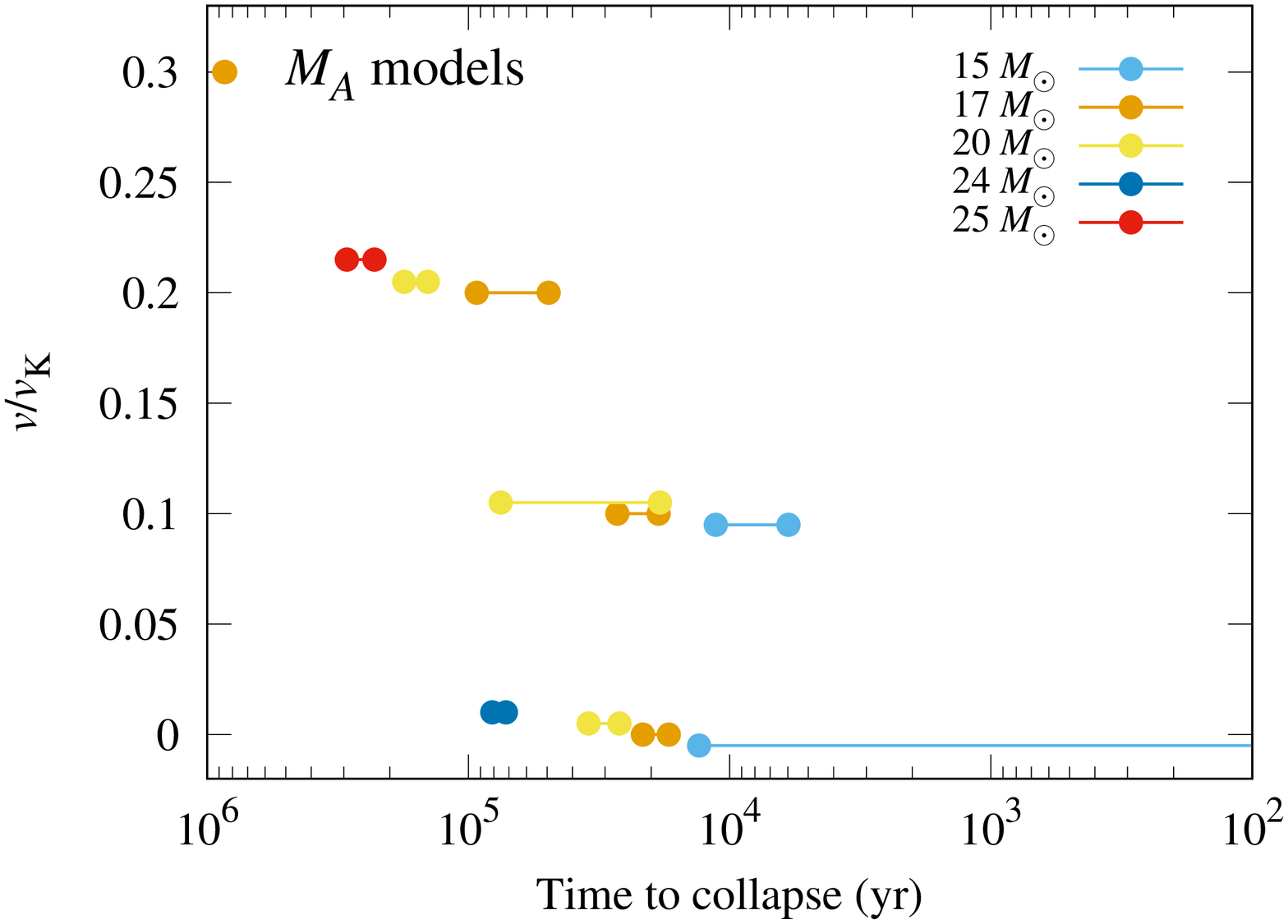}
    \caption{The timing of the fit for certain models with varying initial parameters, shown as a comparison.}
    \label{fig:timingoffit}
\end{figure}

Betelgeuse's current stage of evolution is still up for debate.
The recent dimming episode has spurred some discussion around this topic, about the possibility that Betelgeuse could undergo an imminent supernova. 
Recent studies mostly label the supernova scenario as conjecture, mostly suggesting the dimming is due to other mechanisms. 
Results from \citet{Dolan2016, Wheeler2017} also limit Betelgeuse to the core helium burning phase.
However, our results somewhat contradict this consensus, as shown in figure \ref{fig:timingoffit}, where we see several models that are able to fit to Betelgeuse until the end of their evolution, suggesting the possibility that Betelgeuse could currently be in or even past the core carbon burning stage. 
In the lower panel of figure \ref{fig:timingoffit}, we can see that most \mov{} overshoot models can only fit for a period of time during core helium burning, and only the \textit{15noM} model can fit until the end of evolution.
This outcome largely echoes the results from \citet{Dolan2016}, who had only considered a lower overshoot parameter, but did not consider 15 \msun{} models in their best fit range.
However, when we consider possibility of the higher \lov{} overshoot models, we are able to find many models of differing initial masses and initial rotation, which can fit to Betelgeuse until the end of its evolution.

The differences in our results also stem from the use of a different set of observational constraints which have not been systematically studied before.
While this does mean that it is rather prone to error, as for example, the error range for surface nitrogen and oxygen abundances are quite large, the additional use of the N/O ratio allows us to put much tighter constraints on our results.
Thus, we believe our results show that the possibility of an imminent supernova event cannot be entirely ruled out.
Further studies to try and reproduce these results, such as using different stellar evolution code suites, or by obtaining more precise measurements of the surface abundances would be worthwhile.

\subsection{Blue Loop Phase} \label{subsec:blueloop_disc}

A small number of our models underwent a blue loop phase during helium burning, which affects our results for the total time of fit, for example the \textit{1702L} model from figure \ref{fig:02rot}.
These models are denoted by a $\dagger$ in table \ref{tab:solsum}.
Blue loop phases are known to be related to a number of input parameters, for example \citet{Walmswell2015} discusses the implications of excess helium around the core and how it causes the blue loop.
It has also been suggested that higher overshoot in the envelope would favour blue loop formation \citep{Ritossa1996}, which would explain why we only find its existence in the \lov{} overshoot models.
However, aside from the overshoot model, there is no discernible trend among our models as to what combination of initial conditions induces this loop phase. 
Both non-rotating and rotating models have blue loop models, and previous simulations using this code (see \citet{Yoshida2019}, figure 12) also resulted in similar loop phases in their 18 \msun{} models. 
Perhaps the occurrence stems from the calculation procedure, and is rather random in nature. 
Nevertheless, during the loop phase, the surface abundances remain largely unchanged, and only the change in HR diagram position affects the fit for Betelgeuse. 
The loop phase occupies the majority of the helium burning phase, which results in shorter a total time of fit when compared to models without the loop phase.
In our \textit{17noL} and \textit{1702L} models, the time spent in the loop outweighs the time spent within Betelgeuse's observed HR diagram error bars by roughly 5 to 3.
However, due to the fact that stellar properties pre- and post-loop phase in our models are similar, we believe the existence of the loop phase and its implications can be largely ignored in the context of its application to Betelgeuse progenitor models.

\subsection{Metallicity} \label{subsec:metallicity_disc}

In this study, we have focused on solar metallicity models, based on the surface abundances inferred from a surface temperature of \SI{3600}{K}.
However, it has been shown that the surface temperature of Betelgeuse is variable.
If we instead consider the surface abundances for a higher surface temperature, such as \SI{3800}{K}, then we would find the constraints from \citet{Lambert1984} to be $\epsilon_c = 8.41 \pm 0.15$, $\epsilon_{\ch{N}} = 8.62 \pm 0.15$, and $\epsilon_{\ch{O}} = 8.77 \pm 0.15$.
Under these constraints, the corresponding [CNO/H] value is 9.10, meaning the solar metallicity values we have used would become unsuitable. 
Instead, we must adjust the metal content to $[Fe/H] \sim$ +0.1 to +0.2 in order to find viable models. 
For reference, on the basis of the \citet{Asplund2009} solar abundances, if we increase the [Fe/H] to +0.1 or +0.2, the total [CNO/H] value becomes 9.06 and 9.17 respectively, and if we consider a middle ground of [Fe/H] = +0.15, the corresponding [CNO/H] value becomes 9.12.

In table \ref{tab:015Zsummary}, we have provided a simple overview of the viability of [Fe/H] = +0.15 using 15, 17, and 20 \msun{} models, using the same observational constraints for everything aside from the surface abundances.
We can see that despite the change in metallicity, the results are qualitatively similar to those for solar metallicity, except for the \textit{1501L} and \textit{15noM} models that are no longer viable.

However, the enhanced metallicity also causes the red giant branch to become cooler.
Therefore, if we also impose different surface temperature constraints, certain models would have the timing of their fit be impacted, or have their viability removed altogether. 
A previous study by \citet{Song2020} suggests that this reduction in surface temperature can be offset by increasing the mixing length parameter $\alpha$, which has been kept at the default value of 1.8 in this study.
We found that an increase of $\alpha$ to $\sim 2.0$ is required to produce the results in table \ref{tab:015Zsummary} for a surface temperature increase of \SI{200}{K}.
Note although the mixing length parameter has been calibrated to $\alpha = 1.8$, it is not necessarily a fixed constant during evolution, and also does not rule out the possibility that Betelgeuse could behave similar to a model with higher mixing length as \citet{Sonoi2019} have shown variation in the calibration of the mixing length parameter.
Overall, we cannot use our results to determine the metallicity of Betelgeuse's progenitor model due to the large uncertainty of the surface temperature. 
The use of the mixing length parameter to make up for surface temperature differences also complicates the matter.

\begin{table*}[t]
    \centering
    \caption{Summary of the fit to Betelgeuse for [Fe/H] = +0.15 models}
    \label{tab:015Zsummary}
    %\resizebox{\columnwidth}{!}{%
    \begin{tabular}{c|c|c|c|c|c|c}
    \tableline
    Rotation &  \multicolumn{2}{c|}{15 \msun}  & \multicolumn{2}{c|}{17 \msun}   &  \multicolumn{2}{c}{20 \msun} \\\cline{2-7}
    ($v/v_{\rm K}$)  & \lov{} & \mov{}  & \lov{} & \mov{} & \lov{} & \mov{} \\
    \tableline
    no-rot  & $\times$ & $\times$ & $\circ$ & $\circ$ & $\circ$ & $\circ$  \\
    \tableline
    0.1     & $\times$ & $\circ$ & $\circ$ & $\circ$  & $\circ$ & $\circ$  \\
    \tableline
    0.2     & $\circ$ & $\times$ & $\circ$ & $\circ$  & $\times$ & $\circ$  \\
    \tableline
    0.4     & $\times$ & $\times$ & $\times$ & $\times$ & $\times$ & $\times$  \\
    \tableline
    \end{tabular}\par
    \bigskip
    {\textbf Notes} -- In this case, the surface abundance constraints are $\epsilon_c = 8.41 \pm 0.15$, $\epsilon_{\ch{N}} = 8.62 \pm 0.15$, and $\epsilon_{\ch{O}} = 8.77 \pm 0.15$, as well as $N/O = -0.15 \pm 0.05$. $\circ$ represents a model with good fit for Betelgeuse on the red giant branch, $\times$ represents a model which does not fit Betelgeuse during its evolution.
\end{table*}

\subsection{Surface Rotation of Betelgeuse} \label{subsec:surfrot_disc}

So far we have neglected an important observational parameter which is Betelgeuse's surface rotation. 
At up to $v \approx$ \SI{15}{km\:s^{-1}} \citep{Kervella2018}, Betelgeuse exhibits abnormally rapid rotation for a RSG, which proved to be troublesome as none of our models were able to sustain high enough surface rotation into the helium burning stage.

The surface rotation of our models sharply drop as the star expands as a supergiant, and regardless of its initial rotation, the observed surface rotation velocity cannot be satisfied.
This result also agrees with the main conclusion from \citet{Wheeler2017}.
In addition, various methods of angular momentum transport from the core to the surface, such as the (absence of) Tayler-Spruit dynamo effect \citep{Heger2000}, \citet{Heger2005} or increasing viscosity \citep{Wheeler2017}, had also been proven to be ineffective at reproducing high surface rotation in RSGs.

There are emerging theories which attempt to explain Betelgeuse's rapid rotation, all of which involve Betelgeuse in a binary system in the past.
One example is a merger theory, suggesting that Betelgeuse had absorbed a smaller companion during its evolution, which spun up its rotation.
Recent studies have shown that a merger between a $\sim$ 15 to 20 \msun{} during the ascent up the red giant branch \citep{Chatzopoulos2020} or during the core helium or core carbon burning stages \citep{Sullivan2020} can allow the model to attain satisfactory surface rotation as a red supergiant.

In our results, we have several models in the 15 to 20 \msun{} range which present a good fit for Betelguese during and after the times of the proposed merger event.
This suggests that if surface conditions were not greatly disturbed by the merger, our models could provide a complete reproduction of Betelgeuese's observed properties.
\citet{Chatzopoulos2020} suggests that surface conditions could change, or we could for see signatures of change during or after the merger, for example if the material mixed into the core regions induces additional mixing and material is brought up to the surface.
However, \citet{Sullivan2020} argues that the inner structure changes in their post-merger models would not necessarily be accompanied by changes on the surface.
In addition, there are scenarios where the merger material is distributed mostly into the outer regions of the primary, which still results in a spin-up but much less impact on the mixing processes, so our results cannot be completely ruled out as invalid in those cases.

Another example involves Betelgeuse being spun-up through accretion of mass from a larger companion over a period of time, and then being ejected by a supernova.
This scenario is briefly discussed in \citet{Chatzopoulos2020}, who referred to it as a possibility but less likely than the merger scenario, and mentioned that mixing of the accreted material deeper into the star could again alter its evolution.
However, this accretion scenario has not been studied in depth, so we cannot draw any concrete conclusions of its impacts on our results.
We believe it is best to think of it similar to the merger scenario, where the most important point of consideration is how far the added material penetrates into the Betelgeuse model, and whether that could lead to significant changes of surface properties.

\section{Summary} \label{sec:summary}

We have tested models of varying initial mass, initial rotation, and overshoot parameters using Betelgeuse's HR diagram position and surface CNO abundances as observational constraints. 
In our results, we found models of initial mass between 12 and 25 \msun{} with a good fit for Betelgeuse. 
Models $\geq$ 15 \msun{} can use either the \mov{} ($f_{ov} = 0.01$) or \lov{} ($f_{ov} = 0.03$) overshoot parameter, while $<$ 15 \msun{} models require the higher \lov{} overshoot parameter to reach satisfactory luminosity. 
With regards to initial rotation, we found that both non-rotating and rotating models are able to produce good fits, but at masses near the lower and upper limits, only rotating models are viable.
Also, above an initial rotation of $v/v_{\rm K} \sim 0.3$, surface conditions as an RSG do not match the observed values.

In some of our models, we found that the model is able to stay a fit for Betelgeuse into core carbon burning or beyond, which differs from previous studies that placed Betelgeuse near the beginning of the red giant branch in core helium burning.
This suggests that Betelgeuse could be closer to the end of its life than previous studies believe, and could undergo a supernova event soon.

Finally, Betelgeuse's current surface rotation remains an unsolved problem.
While we have found promising candidates within our grid of models that can conform to both merger theories from \citet{Chatzopoulos2020} and \citet{Sullivan2020}, the exact details of the post-merger models are still unclear. 
Complications arise when considering the surface composition of the post-merger model if the added material is mixed deep into core regions of the primary.
Such dramatic changes to the surface conditions would affect our results, so a comprehensive study on post-merger model properties compared with observed values would be a logical next step.
However, there is also the possibility that added material is mostly added to the outer envelope regions, in which case our results could provide a complete reproduction of Betelgeuse's observed properties.

Overall, we consider our results as contrary to previous studies, opening up new possibilities and discussions for Betelgeuse's current and past evolution.
In future studies, we hope to find tighter limits on the progenitor model using more precise observational constraints in surface abundance and temperature, as well as better implementation of overshoot parameters.
Investigating the merger scenarios and the post-merger behaviour with respect to those constraints will also hopefully shed light on their viability as a method to produce rapidly rotating RSGs like Betelgeuse.

\acknowledgments

This study was supported in part by the Grants-in-Aid for the Scientific Research of Japan Society for the Promotion of Science (JSPS) KAKENHI Grant Number (JP17K05380,JP20H05249,JP20H00158,JP21H01123).

% %% For this sample we use BibTeX plus aasjournals.bst to generate the
% %% the bibliography. The sample63.bib file was populated from ADS. To
% %% get the citations to show in the compiled file do the following:
% %%
% %% pdflatex sample63.tex
% %% bibtext sample63
% %% pdflatex sample63.tex
% %% pdflatex sample63.tex

\bibliography{main}{}
\bibliographystyle{aasjournal}

\end{document}